\documentclass{elsart}
\usepackage{amsmath}
\usepackage{amssymb}
\usepackage{times}
\usepackage{graphicx}
\newsymbol\lessim 132E
\begin{document}

\begin{frontmatter}

\title{An exactly solvable nonlinear model: Constructive effects of
correlations between Gaussian noises}

\author{P.  F.  G\'ora}
\ead{gora@if.uj.edu.pl}
\address{M.~Smoluchowski Institute of Physics and Mark Kac Complex Systems 
Research Center, Jagellonian University, Reymonta~4, 30--059~Krak\'ow, Poland}

\begin{abstract}
A system with two correlated Gaussian white noises is analysed. This system can
describe both stochastic localization and long tails in the stationary distribution.
Correlations between the noises
can lead to a nonmonotonic behaviour of the variance as function of the intensity
of one of the noises and to a stochastic resonance. A method for improving the 
transmission of external periodic signal by tuning parameters of the system
discussed in this paper is proposed.
\end{abstract}


\end{frontmatter}

\section{Introduction}

Stochastic models with multiplicative, or parametrical, noise find numerous applications
in a variety of branches of science and technology. Unfortunately, models for which
analytical results are known are very scarce and any such a model deserves a thorough
discussion. Recently, Denisov and Horsthemke \cite{denisov1} have discussed a model
given by the equation

\begin{equation}\label{vitr:denisov}
\dot x = -ax + |x|^\alpha \eta(t)\,,
\end{equation}

\noindent where $0\leqslant\alpha\leqslant1$, $\eta(t)$ is a Gaussian noise, possibly 
coloured, and have found that it can describe anomalous diffusion and stochastic 
localization. Denisov and Horsthemke have also discussed several physical systems
in which models of the type \eqref{vitr:denisov} can be useful; see references
provided in their paper. Later Vitrenko~\cite{vitrenko} has generalized \eqref{vitr:denisov} 
to include two noise terms:

\begin{equation}\label{vitr:vitrenko}
\dot x = -(a + \eta_1(t))x + |x|^\alpha \eta_2(t)\,,
\end{equation}

\noindent where $\eta_{1,2}$ are certain coloured and correlated Gaussian noises.
This system has a very nice feature: for $0<\alpha<1$, it
interpolates between a linear transmitter with multiplicative and additive noises 
($\alpha=0$) and a linear system with a purely multiplicative noise ($\alpha=1$). 
The two limiting cases, $\alpha=0$ and $\alpha=1$, are very well known in the 
literature (see e.g.\ Ref.\cite{physa} and references quoted therein).
Vitrenko has formally linearized the system \eqref{vitr:vitrenko} by means of 
a~substitution that has been already used in~\cite{denisov1}:

\begin{equation}\label{vitr:substitution} 
y = \frac{x}{|x|^\alpha}
\end{equation}

\noindent and solved the resulting equation for the trajectories. Converting back to
the original variable proves to be rather tricky and that author has managed to do 
so only if the noises $\eta_{1,2}$ are correlated in a very specific (not to say 
peculiar) manner. It is now widely recognized that correlations between 
various noises can lead to many interesting effects. It is, however, possible that 
phenomena reported by Vitrenko result principally from the very specific form of
correlations assumed by this author and are not generic to the 
system~\eqref{vitr:vitrenko}. We find it interesting to see how the system behaves 
for the intermediate values of $\alpha$ when the correlation requirements are 
less restrictive than those discussed by Vitrenko.

Coloured noises introduce more complexity. However, if 
a dynamical effect is present in the white noise case, it also appears, perhaps in
a distorted form, in the coloured case \cite{peter}. To simplify the discussion,
we will assume that the noises are white. Finally, note that the expression 
$a +\eta_1(t)$ in Eq.~\eqref{vitr:vitrenko} can be interpreted as a~biased noise. 
The noise that multiplies $|x|^\alpha$ in Eq.~\eqref{vitr:vitrenko} is not biased. 
To ``symmeterize'' the system, we include a bias in $\xi_2$ in our analysis. 
It is also convenient to have explicit expressions for noise amplitudes, or 
coupling constants between the noises and the dynamical variable. We thus recast 
the equation \eqref{vitr:vitrenko} in the form

\begin{equation}\label{vitr:langevin}
\dot x = - (a + p\,\xi_1(t))x + |x|^\alpha(b + q\,\xi_2(t))\,,
\end{equation}

\noindent where $a>0$, $0\leqslant\alpha\leqslant1$, $\xi_{1,2}$ are 
mutually correlated Gaussian white noises:

\begin{subequations}\label{GWN}
\begin{gather}
\left\langle\xi_i(t)\right\rangle=0\,,\quad
\left\langle\xi_i(t)\xi_i(t^\prime)\right\rangle=\delta(t-t^\prime)\,,\quad i=1,2\,,\\
\left\langle\xi_1(t)\xi_2(t^\prime)\right\rangle=c\,\delta(t-t^\prime)
\end{gather}
\end{subequations}

\noindent and $c\in[-1,1]$. If not otherwise specified, we interpret the noises 
in the sense of Ito. For the sake of terminology, we will call the noise
$\xi_1(t)$ ``multiplicative'' and $\xi_2(t)$ ``additive'', even though this terminology
is accurate only if $\alpha=0$ (for $\alpha>0$ both noises couple parametrically).
Note that if a particle hits $x=0$, it stays there forever if $\alpha>0$.

There is, in fact, one more reason for including $b\not=0$ in our discussion.
Much as the substitution \eqref{vitr:substitution} linearizes the system
\eqref{vitr:langevin}, another substitution, namely

\begin{equation}\label{vitr:another}
z = \frac{|x|^\alpha}{x}
\end{equation}

\noindent converts it to a noisy logistic equation

\begin{equation}\label{vitr:logistic}
\dot z = (1-\alpha)(a+p\,\xi_1(t))z - (1-\alpha)(b + q\,\xi_2(t))z^2\,.
\end{equation}

\noindent We have discussed this last system in \cite{acta2} and found that 
$b\not=0$ together with correlations between the noises can lead to 
a~nonmonotonic behaviour of the variance $\left\langle z^2\right\rangle - 
\left\langle z\right\rangle^2$ as a function of the intensity of the 
``additive'' noise, $q$, and to a stochastic resonance \cite{SRreview} if 
the system is additionally stimulated by an external periodic signal. 
It would be na\"\i{}ve to expect that these phenomena occur in the system
\eqref{vitr:langevin} in exactly the same manner as they do in \eqref{vitr:logistic}.
A~nonlinear change of variables, especially in case of stochastic equations,
can significantly alter the behaviour. We will see, however, that there are
striking similarities between the systems~\eqref{vitr:logistic} and~\eqref{vitr:langevin}.

This paper is organized as follows: We construct the Fokker-Planck equation
for the system \eqref{vitr:langevin} in Section~\ref{FP} and in 
Section~\ref{distributions} we present its stationary solutions. Then in 
Section~\ref{constructive} we discuss the constructive effects of the
correlations between the noises; in particular, in Section~\ref{SR}
we give numerical evidence for the presence of the stochastic resonance.
Conclusions are given in Section~\ref{conclusions}.

\section{The Fokker-Planck equation}\label{FP}

The problem of constructing a Fokker-Planck equation corresponding to 
a~process driven by two correlated Gaussian white noises has been first
discussed in Ref.~\cite{Telejko}, where the two noises have been decomposed
into two independent processes. The same result has been later re-derived
in \cite{cao94}, where the authors have attempted to avoid an explicit
decomposition of the noises but eventually resorted to a disguised form of
the decomposition. The Fokker-Planck equation for correlated white noises 
has been also discussed in Refs.~\cite{duffing} and~\cite{denisov2} 
and in several other papers; see, for example, Ref.~\cite{acta2} for 
a~particularly simple re-derivation. 

A general Langevin equation

\begin{equation}\label{fp:general}
\dot x = h(x) + g_1(x)\xi_1(t) + g_2(x)\xi_2(t)\,,
\end{equation}

\noindent where $x(t)$ is a one-dimensional process and $\xi_{1,2}$ are 
as in Eqns.~\eqref{GWN}, leads to the following Fokker-Planck 
equation in the Ito interpretation:

\begin{subequations}\label{fp:general-FP}
\begin{equation}
\frac{\partial P(x,t)}{\partial t} =
-\frac{\partial}{\partial x}h(x)P(x,t) +
\frac{1}{2}\frac{\partial^2}{\partial x^2}B(x)P(x,t)\,,
\end{equation}
\noindent where
\begin{equation}
B(x) = [g_1(x)]^2 + 2c\,g_1(x)g_2(x) + [g_2(x)]^2\,.
\end{equation}
\end{subequations}

\noindent In case of Eq.~\eqref{vitr:langevin} we obtain

\begin{equation}\label{vitr:Fokker}
\frac{\partial P(x,t)}{\partial t} = 
\frac{\partial}{\partial x}(ax -b|x|^\alpha)P(x,t)
+\frac{1}{2}\frac{\partial^2}{\partial x^2}
\left(p^2x^2 - 2cpqx|x|^{\alpha} + q^2|x|^{2\alpha}\right)P(x,t)\,.
\end{equation}

\noindent In the following we interpret $|x|^\alpha$ as 
$|x|^\alpha = \left(x^2\right)^{\alpha/2}$, where the square of $x$ must be 
calculated prior to taking the fractional power. It is also apparent that the
probability $P(x,t)$ does not depend on the absolute signs of the amplitudes
$p$, $q$, but only on their relative sign. We assume that $\text{sgn}(pq)=+1$.
This comes at no loss of generality as the equation~\eqref{vitr:Fokker} is 
invariant under a~simultaneous change of signs of $pq$ and~$c$.  

Finding stationary distributions corresponding to Eq.~\eqref{vitr:Fokker}
is the main goal of this paper. This, in principle, could be handled by standard
methods \cite{Risken}, but it would be very difficult due to the absolute value
and the fractional powers. It is apparent that since the right-hand-side of
the corresponding stationary equation vanishes identically if $x=0$, the
term $\delta(x)$ should always be included in any stationary distribution.
We now use the substitution \eqref{vitr:substitution}. After some algebra 
we eventually obtain

\begin{eqnarray}
\frac{\partial P(y,t)}{\partial t} &=& (1-\alpha)\frac{\partial}{\partial y}
\left[a y - b + \frac{\alpha}{2y}(p^2y^2 -2cpqy + q^2)\right] P(y,t)
\nonumber\\\label{vitr:Fokker-new}
&+&
\frac{1}{2}(1-\alpha)^2\frac{\partial^2}{\partial y^2}
(p^2y^2 -2cpqy + q^2) P(y,t)\,.
\end{eqnarray}

\noindent The last term in the square brackets in Eq.~\eqref{vitr:Fokker-new} 
corresponds to the Ito interpretation~\cite{vanKampen}. This term is missing 
if the noises are interpreted according to Stra\-tonovich.

The stationary distribution solves an equation that is fairly easy to integrate:

\begin{gather}
\left[\left(a +\left(1-\frac{1}{2}\alpha\right)p^2\right) y - b
- cpq + \frac{\alpha q^2}{2y}\right]P_{\text{st}}
\nonumber\\\label{vitr:Fokker-stationary}
{}+\frac{1}{2}(1-\alpha)(p^2y^2-2cpqy+q^2)\frac{dP_{\text{st}}}{dy} = 0\,.
\end{gather}

\section{Stationary distributions}\label{distributions}

Before proceeding to the general case, let us discuss the case where there is only
one noise present. 

\subsection{No ``multiplicative'' noise}

If there is no ``multiplicative'' noise, $p=0$, and no bias in the ``additive'' 
noise, $b=0$, our problem reduces to that discussed in Ref.~\cite{denisov1}.
Eq.~\eqref{vitr:Fokker-stationary} takes the form

\begin{equation}\label{vitr:reduced1}
\left(a y + \frac{\alpha q^2}{2y}\right)P_{\text{st}} 
+ \frac{1}{2}(1-\alpha)q^2\frac{dP_{\text{st}}}{dy}=0\,.
\end{equation}

\noindent This equation corresponds to the following Langevin equation

\begin{equation}\label{vitr:reduced2}
\dot y = -\left(a y + \frac{\alpha q^2}{2y}\right) + \sqrt{1-\alpha}q\,\xi_2(t)\,,
\end{equation}

\noindent which, in turn, corresponds to an overdamped motion in a potential

\begin{equation}\label{vitr:reduced-potential}
V_{\text{eff}}(y) = \frac{1}{2}a y^2 + \frac{1}{2}\alpha q^2 \ln|y|\,.
\end{equation}

\noindent The effective potential \eqref{vitr:reduced-potential} has an infinite
noise-created well at $y=0$ which traps Brownian particles; this well is missing 
if the noises are
interpreted according to Stratonovich. Curiously, in another context 
we have observed a similar phenomenon, where noise interpreted according to Ito 
created an insurmountable barrier restricting particles to one half of the real 
axis~\cite{njp}. A similar barrier is observed in the noisy logistic system
\eqref{vitr:logistic}, cf.~Ref.~\cite{acta2}. Loosely speaking, the change of 
variables \eqref{vitr:another} converts an infinite barrier into an infinite well. 

The equation \eqref{vitr:reduced1} is solved by

\begin{equation}\label{vitr:reduced-solution-y}
P_{\text{st}}(y) = \frac{N}{\left\vert y\right\vert^{\alpha/(1-\alpha)}}
\exp\left(-\frac{ay^2}{(1-\alpha)q^2}\right),
\end{equation}

\noindent where $N$ is a normalization constant, or transforming back to the original 
variable

\begin{equation}\label{vitr:reduced-solution-x}
P_{\text{st}}(x) = \frac{N}{\left\vert x\right\vert^\alpha}
\exp\left(-\frac{a\left(x^2\right)^{1-\alpha}}{(1-\alpha)q^2}\right).
\end{equation}

The distribution \eqref{vitr:reduced-solution-x} is normalizable for $a>0$ and 
$0\leqslant\alpha<1$. For $\alpha=0$ it reduces to a standard Gaussian distribution, 
and for\ $0<\alpha<1$ it mildly diverges at $x=0$, where some particles are 
trapped, or stochastically localized. If $\alpha$ increases towards unity, the 
divergence becomes more pronounced, or more particles become localized. At the
same time, though, tails of the distribution get heavier which is characteristic 
for anomalous diffusion. If we interpret the system \eqref{vitr:langevin} with 
$b=0$ as ``interpolating'' between linear systems with an additive 
and multiplicative noises, we can see that if $\alpha=0$, the 
stationary distribution is non-singular. The ``less additive'' the system 
becomes as $\alpha$ increases, the more pronounced the singularity is and the tails
of the distribution get flatter. Eventually, for a linear and purely multiplicative
system, the distribution reduces to $\delta(x)$ as all particles collapse to
the origin of the force.

The presence of a bias, $b\not=0$, introduces some asymmetry in the exponential
term, but the overall behaviour remains much the same:

\begin{equation}\label{vitr:asymmetry}
P_{\text{st}}(x) = \frac{N}{\left\vert x\right\vert^\alpha}
\exp\left(\frac{2bx|x|^{-\alpha} -a\left(x^2\right)^{1-\alpha}}{(1-\alpha)q^2}\right).
\end{equation}

If the noises are interpreted according to Stratonovich, we obtain

\begin{equation}\label{vitr:reduced-stratonovich}
P_{\text{st}}^{\text{Strat}}(x) = N^\prime
\exp\left(\frac{2bx|x|^{-\alpha} -a\left(x^2\right)^{1-\alpha}}{(1-\alpha)q^2}\right)
\end{equation}

\noindent and there is no stochastic localization.

\subsection{No ``additive'' noise}

If there is no ``additive'' noise, $q=0$, and no bias, $b=0$, the only normalizable
stationary solution is $P_{\text{st}}(x) = \delta(x)$, corresponding to all particles
eventually collapsing to their common resting point. If $b\not=0$, there is no
stationary solution as some particles go to the resting point, but some can escape
to infinity.

\subsection{The general case}\label{general}

If the noises are \textit{not} maximally correlated, $|c|\not=1$, the general solution
reads

\begin{equation}\label{vitr:general}
P_{\text{st}}(x) = \frac{N\,\exp\left[\frac{2(bp-caq)}{\sqrt{1-c^2}\,p^2q}\,
\arctan\left(\frac{px|x|^{-\alpha} -cq}{\sqrt{1-c^2}\,q}
\right)\right]}
{|x|^\alpha
\left[
q^2 - 2cpq x |x|^{-\alpha} + p^2 (x^2)^{1-\alpha}
\right]^{\left(1+\frac{a}{(1-\alpha)p^2}\right)}}\,,
\end{equation}

\noindent where $N$ is again a normalization constant. Despite its complicated form,
principal properties of the distribution \eqref{vitr:general} are easy to find.
Because the inverse tangens function, $\arctan(\cdot)$, is limited, the
exponential term is also limited and convergence properties of \eqref{vitr:general}
depend solely on its denominator. One can easily see that this distribution is
normalizable for all $0\leqslant\alpha<1$. For $0<\alpha<1$ the stochastic localization,
in the Ito interpretation, occurs. The distribution has rather heavy
tails. It has a~convergent first moment if $a>\frac{1}{2}\alpha p^2$. The second moment 
is convergent if a~stronger condition, $a>\frac{1}{2}(1+\alpha)p^2$, is satisfied.

If either $b\not=0$, or $c\not=0$, or both, the distribution \eqref{vitr:general}
is not symmetric. Apart from the narrow singularity around $x=0$, it has
another peak centred on the minimum of
$q^2 - 2cpq x |x|^{-\alpha} + p^2 (x^2)^{1-\alpha}$. Its location depends on the 
sign of $c$: if $c>0$, the peak is located to the right of $x=0$, and if $c<0$, 
it is located to the left. If $|c|\lessim1$, the height of this peak can be very 
large. Thus, in the stationary state, a~majority of Brownian particles is 
stochastically localized either around the central singularity, or in the 
additional peak created by the correlations. The tails do not contribute much 
to the overall density. However, if $\alpha$ approaches unity, the tails, and 
the tail with the same sign as the location of the peak in particular, become 
rather heavy and outliers, or particles far removed from both the peak and the 
singularity, can easily be found. The asymmetry between the tails is introduced 
by the exponential term: The distribution is further asymmetrically broadened by 
the exponential term in the numerator of~\eqref{vitr:general}. This broadening can 
be removed if 

\begin{equation}\label{vitr:condition-c}
bp-caq=0\,.
\end{equation}

\noindent It is important to understand the origin of this phenomenon. The 
asymmetric broadening results from the bias --- the force acting in one direction
is, on the average, larger than the force acting in the opposite one. In the 
system \eqref{vitr:langevin} the parameter $b\not=0$ acts as one source of
the bias; it has been introduced for this specific purpose.
It is also known that correlations between two noises can effectively introduce
another bias, see e.g.\ Refs.~\cite{physa,Telejko,duffing}. If the condition
\eqref{vitr:condition-c} is met, the two sources of bias nullify each other.
To see this, let us represent the two correlated Gaussian white noises 
$\xi_{1,2}$ as linear combinations of two \textit{independent} GWNs $\psi_{1,2}$:

\begin{subequations}\label{cholesky}
\begin{eqnarray}
\xi_1(t) & = & \phantom{c}\,\psi_1(t)\,,\\
\xi_2(t) & = & c\,\psi_1(t) + \sqrt{1-c^2}\,\psi_2(t)\,.
\end{eqnarray}
\end{subequations}

\noindent With the condition \eqref{vitr:condition-c} satisfied,
the Langevin equation \eqref{vitr:langevin} now takes the form

\begin{equation}\label{vitr:special}
\dot x = -(a + p\psi_1(t))\left(x-\frac{b}{a}|x|^\alpha\right)
+\sqrt{1-c^2}\,q\,|x|^\alpha \psi_2(t)\,.
\end{equation}

\noindent The system now behaves as if it were driven by two uncorrelated
white noises, one of which is unbiased. As a result, the bias-induced
asymmetric broadening of the stationary distribution disappears.

\subsection{Maximally correlated noises}

The distribution \eqref{vitr:general} does not have a universal limit $|c|\to1$.
Instead, if $c=\pm1$, we need to solve Eq.~\eqref{vitr:Fokker-stationary} directly 
and then convert back to the original variable. We obtain a~candidate solution

\begin{equation}\label{vitr:trial}
P_{\text{trial}}(x) \sim
\frac{\exp\left(\frac{2(bp\mp aq)}{(1-\alpha)p^2(q \mp px |x|^{-\alpha})}\right)}
{|x|^\alpha 
\left\vert\vrule width0pt depth5pt
q \mp px |x|^{-\alpha}\right\vert^{2\left(1+\frac{a}{(1-\alpha)p^2}\right)}}\,,
\end{equation}

\noindent where the $\mp$ sign is the opposite of the sign of the correlation
coefficient, $c=\pm1$. However, the right-hand-side of \eqref{vitr:trial} 
is not normalizable.
If $q \mp px |x|^{-\alpha}=0$, the exponential in \eqref{vitr:trial} hits its 
essential singularity. This singularity is eliminated if a~special case
of the condition \eqref{vitr:condition-c}, namely

\begin{equation}\label{vitr:condition}
bp\mp aq=0\,,
\end{equation}

\noindent holds. In this case, either $p=q=0$ and the system becomes fully 
deterministic, or the stationary Fokker-Planck equation 
\eqref{vitr:Fokker-stationary} factorises:

\begin{equation}\label{vitr:factorize}
\frac{py \mp q}{2py}
\left[
[(2a + (2-\alpha)p^2)y \mp\alpha p q]P_{\text{st}}(y) 
+ (1-\alpha)py(py\mp q)\frac{dP_{\text{st}}(y)}{dy}
\right]=0\,.
\end{equation}

\noindent A singular distribution $\delta(py\mp q)$ solves Eq.~\eqref{vitr:factorize}.
The regular part of this equation, the one in the square brackets, again leads to a not
normalizable solution. We, therefore, conclude that if the noises are maximally
correlated and the condition \eqref{vitr:condition} holds, the stationary
distribution reads

\begin{equation}\label{vitr:twodeltas}
P_{\text{st}}(x) = \gamma\delta(x) + (1-\gamma)\delta\left(x\mp(q/p)^{1/(1-\alpha)}\right),
\end{equation}

\noindent where $\gamma$ is the fraction of the initial population that collapses
to zero; observe that with our sign convention adopted, $q/p>0$. If the noises
are maximally correlated but the condition \eqref{vitr:condition} is not satisfied,
there is no stationary distribution. There is a striking similarity between the
system \eqref{vitr:langevin} and \eqref{vitr:logistic}, where a similar situation
occurs~\cite{acta2}: If a condition analogous to \eqref{vitr:condition} is satisfied
and the noises are maximally correlated, the noisy logistic system has a $\delta$-like 
stationary distribution. If the noises are maximally correlated but the counterpart 
of the condition \eqref{vitr:condition} does not hold, a~stationary distribution does 
not form in the noisy logistic system, either.

\section{Constructive effects of correlations}\label{constructive}

As we have seen, a delicate interplay between the correlations and the bias can
significantly alter the shape of the stationary distribution. We may expect that
this can lead to various unexpected properties of the system \eqref{vitr:langevin}.

\begin{figure}
\begin{center}
\includegraphics{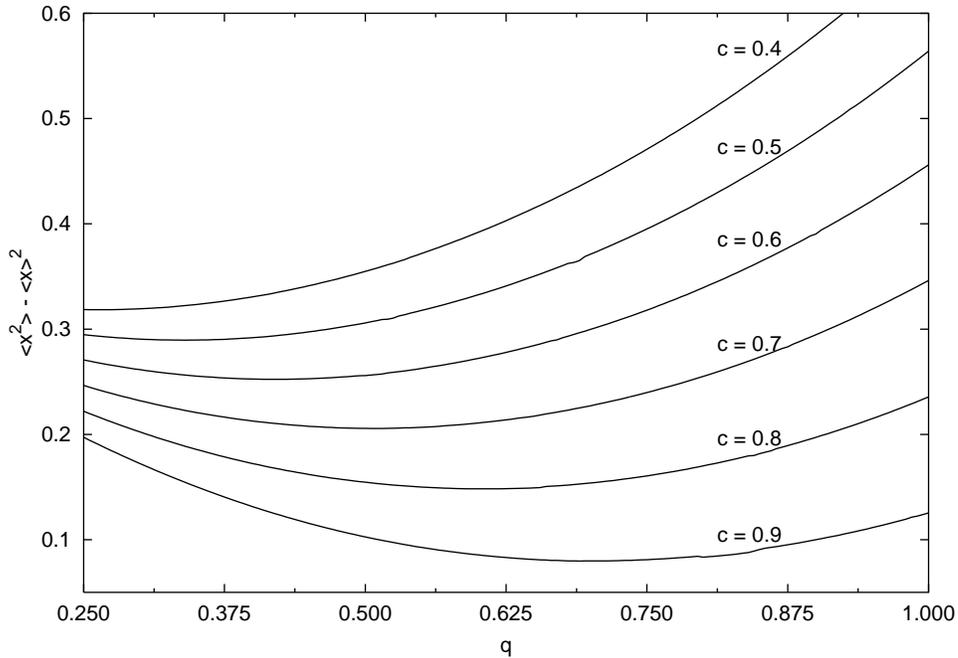}
\end{center}
\caption{Nonmonotonic behaviour of the variance of the distribution \eqref{vitr:general}
as a function of the ``additive'' noise strength, $q$, for various correlations
between the noises, $c$. Other parameters are $\alpha=1/8$, $a=1.25$, 
$b=1.0$, $p=1.0$.}
\label{Fig-variance}
\end{figure}

\subsection{Nonmonotonic behaviour of the variance}

Recall that depending on the parameters, the stationary distribution of the system
discussed in this paper can be nearly limited to very narrow peaks; with a different
set of parameters, these peaks can be asymmetrically broadened. The second central 
moment of a~probability distribution, $\left\langle x^2\right\rangle - 
\left\langle x\right\rangle$, if convergent, is perhaps one of its simplest and 
most easily comprehended characteristics. It is interesting to see how the second
moment of the distribution \eqref{vitr:general} behaves as a~function of the
``additive'' noise strength. Because of the complicated analytical structure
of this distribution, we have not been able to evaluate the integrals
$\int_{-\infty}^\infty x\,P_{\text{st}}(x)\,dx$, 
$\int_{-\infty}^\infty x^2\,P_{\text{st}}(x)\,dx$ analytically. We have done
so numerically instead. Example results are presented in Fig.~\ref{Fig-variance};
parameters chosen correspond to a convergent second moment.
As we can see, a~minimum of the variance as a~function of the ``additive''
noise strength is clearly visible. This minimum is fairly deep if the 
correlations are large and becomes very shallow as the correlations 
decrease. Note that if $b<0$, the minimum appears for negative values
of the correlation coefficient (not plotted).

\subsection{Stochastic resonance}\label{SR}

\begin{figure}
\begin{center}
\includegraphics{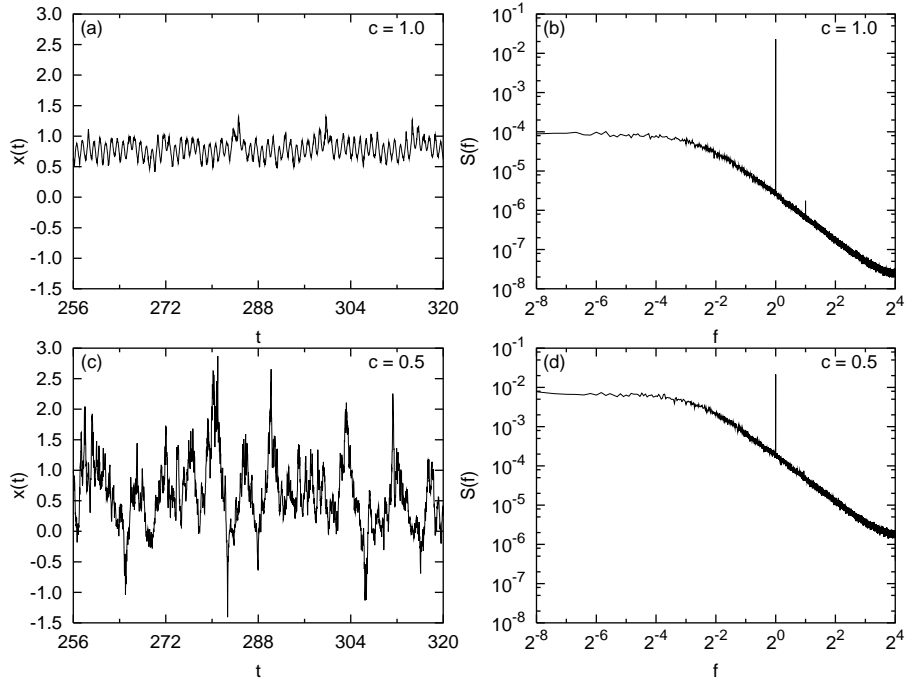}
\end{center}
\caption{Panel (a): a fragment of a typical realization of the process 
\eqref{vitr:forcing} with $c=1$. Panel (b): the corresponding power spectrum 
averaged over 128 realizations. Panels (c), (d): same as (a), (b) above, but 
with $c=0.5$. Other parameters, common for all panels: $\alpha=1/8$, $a=1.25$, 
$b=1.0$, $p=1.0$, $q=0.8$, $A=1$, and $\Omega=2\pi$.}
\label{Fig-runs}
\end{figure}

Now suppose that the system discussed in this paper is additionally stimulated by
an external, periodic signal. The Langevin equation takes the form
\begin{equation}\label{vitr:forcing}
\dot x = - (a + p\,\xi_1(t))x + |x|^\alpha(b + q\,\xi_2(t) + A\cos(\Omega t+\phi))
\end{equation}
\noindent where the noises are as in \eqref{GWN}. Because we do not know exact solutions of 
a~time-dependent Fokker-Planck equation corresponding to 
Eq.~\eqref{vitr:forcing}, we have solved the equation \eqref{vitr:forcing} numerically
with the Euler-Maryuama algorithm and a timestep equal $2^{-12}$. To generate 
the correlated noises $\xi_{1,2}$, we have first generated two independent Gaussian 
white noises $\psi_{1,2}$; we have used the Marsaglia algorithm \cite{Marsaglia} for
that purpose and the famous Mersenne Twister \cite{Mersenne} has been used as 
the underlying uniform generator. Then the correlated noises are created as
linear combinations of the two uncorrelated ones, see Eq.~\eqref{cholesky} above.
Example trajectories of the system \eqref{vitr:forcing} and associated power spectra, 
averaged over 128 realizations
of the noise and on the initial phase of the signal, $\phi$, are presented 
in Fig.~\ref{Fig-runs}. The shape of the trajectories and the power spectra
strongly depend on the parameters of the system, and on the correlation coefficient,
$c$, and the strength of the ``additive'' noise, $q$, in particular. In general,
the higher the correlations, the more ordered the trajectories are. It is worth
noting that higher harmonics of the driving frequency can be visible in the
power spectra, indicating a~nonlinear nature of the coupling between the signal
and the dynamical variable.

\begin{figure}
\begin{center}
\includegraphics{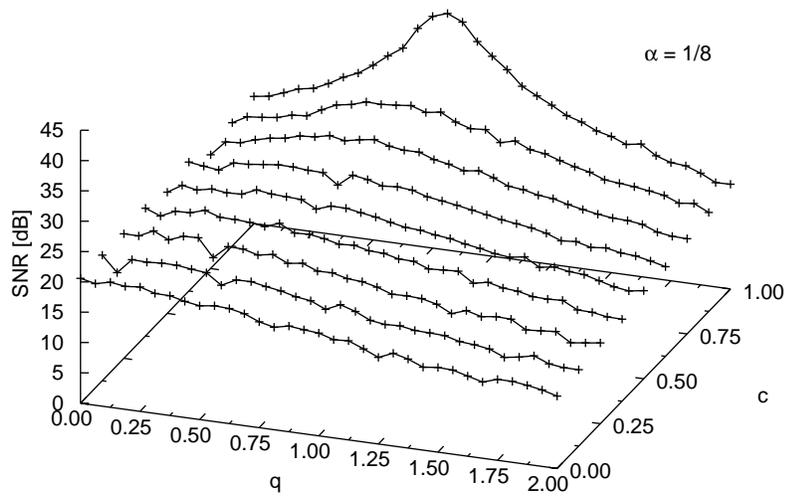}
\end{center}
\caption{Signal-to-noise ratio for the system \eqref{vitr:forcing}. Parameters 
are $\alpha=1/8$, $a=1.25$, $b=1.0$, $p=1.0$, $A=1$, and $\Omega=2\pi$.
Curves presented correspond, back to front, to the following values of the
correlation coefficient: $c=1.000$, $0.875$, $0.750$, $0.625$, $0.500$, $0.375$,
$0.250$, $0.125$, and $0.000$.}
\label{Fig-snr1}
\end{figure}

\begin{figure}
\begin{center}
\includegraphics{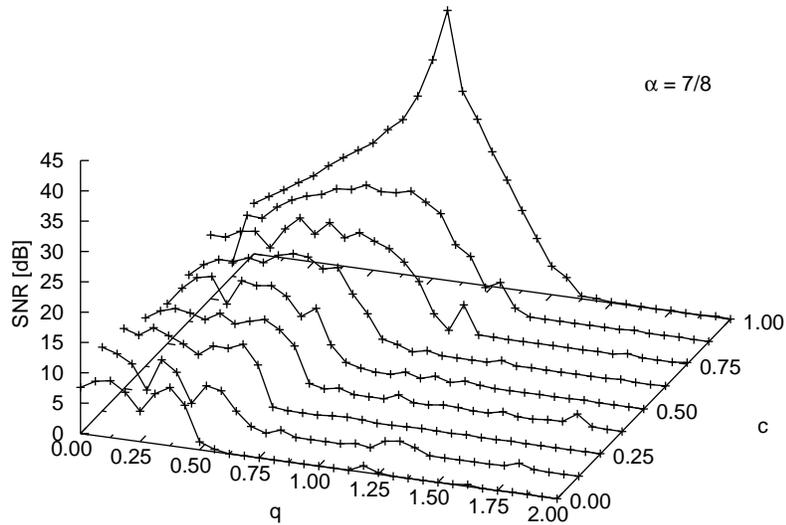}
\end{center}
\caption{Same as Fig.~\ref{Fig-snr1} but with $\alpha=7/8$.}
\label{Fig-snr2}
\end{figure}

To quantify these observations, we will use the Signal-To-Noise Ratio (SNR) as 
a~measure of the stochastic resonance:
\begin{equation}\label{appb:SNR-def}
\text{SNR} = 
10 \log_{10}\frac{S_{\text{signal}}}{S_{\text{noise}}(f=\Omega/2\pi)}\,,
\end{equation}
\noindent where $S_{\text{signal}}$ is the height of the peak in the 
power spectrum at the driving frequency and $S_{\text{noise}}(f)$ is the 
frequency-dependent noise-induced background. Several other measures of the 
stochastic resonance have
been proposed~\cite{spectralamplification}, but we choose the SNR as the
simplest, oldest and most commonly used one.
Selected results, averaged on both realizations of the noises and
the initial phase, are presented in Figs.~\ref{Fig-snr1}
and~\ref{Fig-snr2}. For high values of the correlation coefficient, a~clear
maximum in the SNR is visible. This shows that there is an optimal level
of the ``additive'' noise that maximizes the ratio of power transmitted 
through coherent oscillations induced by the driving signal to that transmitted 
by the irregular ones, or that there is a stochastic resonance in the 
system~\eqref{vitr:forcing}. For correlations only slightly larger than zero,
the resonance is very small and it disappears for $c\leqslant0$. Note that this
happens if the asymmetry parameter, $b$, is greater than zero. For $b<0$
the stochastic resonance occurs for negative correlations and reaches its
largest magnitude at $c=-1$. In the symmetric case, $b=0$, there is no stochastic
resonance. Again, these features of the stochastic resonance resemble very
much those of the noisy logistic system \eqref{vitr:logistic} discussed
in~\cite{acta2}.

The resonance becomes sharper if $\alpha$ approaches unity (Fig.~\ref{Fig-snr2}).
At the same time, values of SNR away from the resonance are much smaller than
those in the small $\alpha$ case.

\subsection{Response to a change of deterministic parameters}

We have shown in the two previous Subsections that the system discussed here
can be optimised by choosing an appropriate level of the ``additive'' noise.
In practice, however, controlling the amplitude of the noise or correlations
between the two sources of the noise can be very difficult. Tuning the
deterministic part of the system may be much easier to achieve, and as
our discussion of the asymmetric broadening of the distribution in 
Subsection~\ref{general} has shown, by changing the bias parameter, $b$,
we can optimise the system even if the noise amplitudes and the correlation
coefficient are not known.

\begin{figure}
\begin{center}
\includegraphics{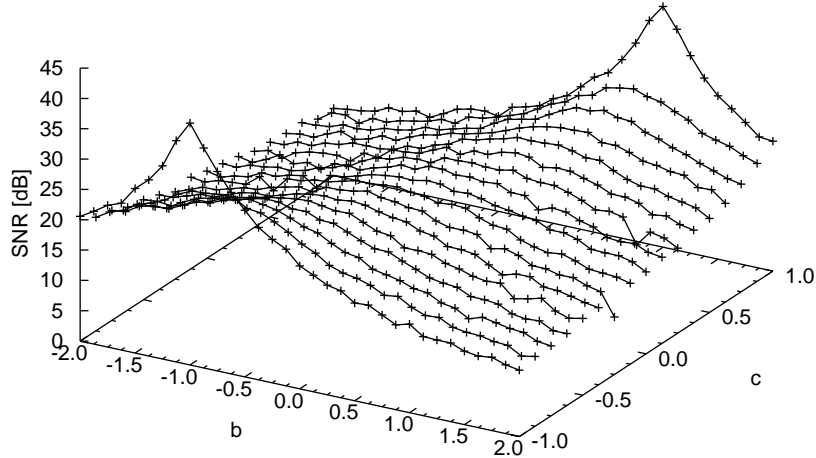}
\end{center}
\caption{Signal-to-noise ratio for the system \eqref{vitr:forcing} as a function
of the bias parameter, $b$, and the correlation coefficient, $c$. Other parameters
are $\alpha=1/8$, $a=1.25$, $p=1.0$, $q=0.8$, $A=1$, and $\Omega=2\pi$.}
\label{Fig-bchange}
\end{figure}

To test for that, we have again numerically simulated the externally 
stimulated system \eqref{vitr:forcing} by the same means that have been
used in Subsection~\ref{SR} above. This time amplitudes of the two noises
have been kept constant and the bias parameter has been varied. Selected 
results are presented in Fig.~\ref{Fig-bchange}. As we can see, changing
the bias does optimise the system. Clear maxima in the signal-to-noise
ratio are visible. These maxima are most pronounced if correlations
are large, $|c|\lessim1$, but they are present also for $|c|\simeq0$,
even though the overall shape of the curves is much flatter. For the 
uncorrelated case, $c=0$, the weak maximum coincides with $b=0$ which is
to be expected due to symmetry of the system. To put it in a slightly
different way, we can see that the uncorrelated system transmits an
external signal badly. Any correlations between the noises potentially
improve the transmission properties. The system can be optimised to reach 
its full potential by appropriately adjusting its deterministic
parameters.

\section{Conclusions}\label{conclusions}

In this paper we have discussed a nonlinear system with two correlated
sources of Gaussian white noises. A closely related system has been discussed 
previously by Vitrenko in Ref.~\cite{vitrenko}. We have been mainly interested 
in what happens when the restrictions on correlations between the noises imposed 
by that author are lifted and, additionally, when the ``additive'' noise becomes 
biased. We have shown that this system can display both stochastic localization
and heavy tails in its stationary distribution which is characteristic for
anomalous diffusion. This agrees with previously published results 
\cite{denisov1,vitrenko}. It is worth noting, though, that authors of that
References obtained their results under the assumption that the noises were
coloured; we have shown that the same happens for white noises as well.

Next, we have shown that correlations present in the system discussed here can
lead to interesting constructive effects of the noise: to a~nonmonotonic behaviour
of the variance of the stationary distribution and to a~stochastic resonance.
Finally, we have shown that the system can be optimised to an external periodic
signal not only by varying amplitudes of the noises, but also by tuning the
deterministic parameters of the system when the noise amplitudes and the
correlation coefficient between the noises remain, in principle, unknown.

Surprisingly, the system \eqref{vitr:langevin} discussed here is related to 
the noisy logistic system \eqref{vitr:logistic} that we have discussed previously 
\cite{acta2}. As we have shown, many, but not all, properties of these two systems 
are strikingly similar.

\end{document}